\documentclass{article}
\usepackage{spconf,amsmath,graphicx,tabto,amsfonts,colortbl,bm, subfig,float,balance}

\title{SPEECH EMOTION RECOGNITION WITH DUAL-SEQUENCE LSTM ARCHITECTURE}
\name{Jianyou Wang$^1$, Michael Xue$^1$, Ryan Culhane$^1$, Enmao Diao$^1$, Jie Ding$^2$, Vahid Tarokh$^1$\thanks{This work was supported in part by Office of Naval Research Grant No. N00014-18-1-2244.}}
\address{$^{1}$ Department of Electrical and Computer Engineering, Duke University, Durham, NC, USA\\
$^{2}$ School of Statistics, University of Minnesota Twin Cities, Minneapolis, MN, USA
}
\begin{document}
\raggedbottom
%\ninept
%
\maketitle
\begin{abstract}
Speech Emotion Recognition (SER) has emerged as a critical component of the next generation of human-machine interfacing technologies. In this work, we propose a new dual-level model that predicts emotions based on both MFCC features and mel-spectrograms produced from raw audio signals. Each utterance is preprocessed into MFCC features and two mel-spectrograms at different time-frequency resolutions. A standard LSTM processes the MFCC features, while a novel LSTM architecture, denoted as Dual-Sequence LSTM (DS-LSTM), processes the two mel-spectrograms simultaneously. The outputs are later averaged to produce a final classification of the utterance. Our proposed model achieves, on average, a weighted accuracy of 72.7\% and an unweighted accuracy of 73.3\%---a 6\% improvement over current state-of-the-art unimodal models---and is comparable with multimodal models that leverage textual information as well as audio signals.    
\end{abstract}

\section{Introduction}
\label{sec:intro}
As the field of Automatic Speech Recognition (ASR) rapidly matures, people are beginning to realize that the information conveyed in speech goes beyond its textual content. Recently, by employing deep learning, researchers have found promising directions within the topic of Speech Emotion Recognition (SER). As one of the most fundamental characteristics that distinguishes intelligent life forms from the rest, emotion is an integral part of our daily conversations. From the broad perspective of general-purposed artificial intelligence, the ability to detect the emotional contents of human speech has far-reaching applications and benefits. Furthermore, the notion that machines can understand and perhaps some day produce emotions can profoundly change the way humans and machines interact.
%There exist several emotion-annotated datasets, most of which consist of human actors repeating the same sentence with different emotions, such as the Berlin Dataset \cite{berlin}. SER models have achieved high accuracy for these datasets. However, in an effort to obtain a more robust model that can detect emotions in daily conversations, researchers have turned their attention to the IEMOCAP corpus \cite{IEMOCAP}, which has proven to be a challenging dataset. For the standard four-emotion classification task on IEMOCAP, state-of-the-art models utilizing only audio data have been unable to achieve an accuracy of 70\% \cite{8682154, iaan}, and multimodal models that utilize both text and audio data have been unable to reach an accuracy of 75\% \cite{multi2, multi1}.

Previous work in SER models on the benchmark IEMOCAP dataset \cite{IEMOCAP} can be generally divided into two categories: unimodal and multimodal. Research that focuses on unimodal data uses only raw audio signals, whereas research in multimodal data leverages both audio signals and lexical information, and in some cases, visual information. Not surprisingly, since they take advantage of more information, multimodal models generally outperform unimodal models by 6-7\%. Traditionally, unimodal models extract high level information from raw audio signals, such as MFCC features, and then pass the output through a recurrent neural network \cite{hand}. Recently, researchers have begun transforming raw audio signals into spectrograms or mel-spectrograms \cite{spec1, spec2}, which contain low level information and can be converted back to raw audio. These spectrograms are then mapped into a latent time series through several convolutional layers before going through a recurrent layer. 

Some researchers think that audio data alone is not enough to make an accurate prediction \cite{multi1}, and thus many have turned to using textual information as well. However, it is possible that two utterances with the same textual content can have entirely different meanings when fueled with different emotions. Therefore, using textual information too liberally may lead to misleading predictions. It is our opinion that the full potential of audio signals has not been fully explored, and we propose several changes to the existing state-of-the-art framework for unimodal SER \cite{8682154, iaan}.

\begin{figure*}[t]%
    \centering
    \includegraphics[width=17cm]{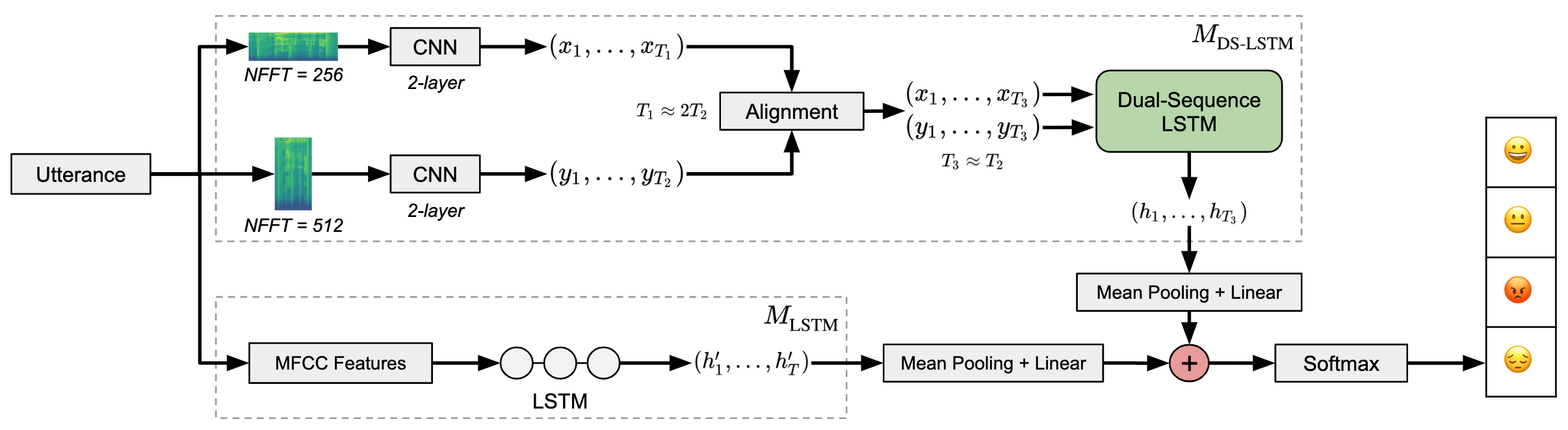}
    \caption{Dual-level model with DS-LSTM cell}
    \label{fig:1}%
\end{figure*}
In this paper, we make three major contributions to the existing unimodal SER framework. First, we propose a new dual-level model that contains two independent neural networks that process the MFCC features and mel-spectrograms separately, but are trained jointly. Similar to other dual-level architectures~\cite{dual-level}, we found that our proposed dual-level model provides a significant increase in accuracy. Second, inspired by the time-frequency trade-off \cite{trade}, from each utterance we calculate two mel-spectrograms of different time-frequency resolutions instead of just one. Since these two spectrograms contain complementary information---namely, one has a better resolution along the time axis and the other has a better resolution along the frequency axis---we propose a novel variant of LSTM \cite{lstm}, denoted as Dual-Sequence LSTM (DS-LSTM), that can process these two sequences of data simultaneously and harness their complementary information effectively. It should be noted that previous research in multi-dimensional LSTM (MD-LSTM)~\cite{multi-dim1,multi-dim2,multi-dim3}, especially in ASR~\cite{ftlstm,gridlstm}, focused on adapting the LSTM to a multi-dimensional data format. Although similar in concept, our proposed DS-LSTM has a distinct architecture, and is designed to process two sequences of one-dimensional data instead of multi-dimensional data. Third, we propose a novel mechanism for data preprocessing that uses nearest-neighbor interpolation to address the problem of variable lengths between different audio signals. We have found that interpolation works better than more typical methods such as truncating and padding data, which lose information and also increase the computational cost.

%The outline of this paper is given next. First, we discuss how we preprocess the data into MFCC features as well as two mel-spectrograms of different time-frequency resolutions. Next, we introduce our proposed dual-level model and describe how it processes these inputs and eventually classifies the utterance. Finally, we outline our experimental procedure and how our model compares to state-of-the-art methods as well as several baseline methods. 
\section{Research Methodology}
\label{sec: methodology}
\subsection{Dataset Description}
\label{ssec: dataset}
We used the Interactive Emotional Dyadic Motion Capture (IEMOCAP) dataset \cite{IEMOCAP} in this work, a benchmark dataset containing about 12 hours of audio and video data, as well as text transcriptions. The dataset contains five sessions, each of which involves two distinct professional actors conversing with one another in both scripted and improvised manners. In this work, we utilize data from both scripted and improvised conversations, as well as only audio data to stay consistent with the vast majority of prior work. We also train and evaluate our model on four emotions: \textit{happy}, \textit{neutral}, \textit{angry}, and \textit{sad}, resulting in a total of 5531 utterances (\textit{happy}: 29.5\%, \textit{neutral}: 30.8\%, \textit{angry}: 19.9\%, \textit{sad}: 19.5\%). We denote these 5531 utterances in the set $\{u_1,\dots,u_{5531}\}$.
\subsection{Preprocessing}
\label{ssec: preprocessing}
For extracting MFCC features, we used the openSMILE toolkit [12], a software that automatically extracts features from an audio signal. Using the MFCC12\_E\_D\_A configuration file, we extracted 13 Mel-Frequency Cepstral Coefficients (MFCCs), as well as 13 delta and 13 acceleration coefficients, for a total of 39 acoustic features. These features are extracted from 25 ms frames, resulting in a sequence of 39-dimensional MFCC features per utterance $u_i \in \{u_1,\dots,u_{5531}\}$.

For each utterance, we also propose to derive two mel-spectrograms of different time-frequency resolutions instead of just one, as done in previous research. One (denoted by $S_{1i}$) is a mel-scaled spectrogram with a narrower window and thus a better time resolution, while the other (denoted by $S_{2i}$) is a mel-scaled spectrogram with a wider window and thus a better frequency resolution. In our work, $S_{1i}$ and $S_{2i}$ are calculated from a short-time Fourier transform with 256 and 512 FFT points, respectively. The hop length and the number of mel channels are 50\% and 25\% of the number of FFT points, respectively. 

The standard method to deal with variable length in utterances is padding or truncation. Since there are rises and cadences in human conversation, we cannot assume the emotional contents are uniformly distributed within each utterance. Therefore, by truncating data, critical information is inevitably lost. On the other hand, padding is computationally expensive. We propose a different approach to deal with variable length between utterances: nearest-neighbor interpolation, in which we interpolate along the time axis for each mel-spectrogram to the median number of time steps for all the spectrograms, followed by a logarithmic transformation.

\subsection{Proposed Model}
\label{ssec: model}
\subsubsection{Dual-Level Architecture}
\label{sssec: joint}
Our proposed dual-level architecture is illustrated in Figure~\ref{fig:1}. It contains two separate models, $M_{\textrm{LSTM}}$ and $M_{\textrm{DS-LSTM}}$, the first for the MFCC features and the second for the two mel-spectrograms. Each of these two models has a classification layer, the outputs of which are averaged to make the final prediction. The loss function is also the average of two different cross entropy losses from the two models.

\subsubsection{LSTM for MFCC Features}
\label{sssec: lstm}
The MFCC features for each utterance are represented by $Z=$ $\{z_1,\dots,z_T\}$, with each $z_i \in \mathbb{R}^{39}$. Each $Z$ is fed into a standard two-layer single-directional LSTM, whose outputs, $H=\{h_1',\dots,h_T'\}$, as specified by Figure~\ref{fig:1}, are mean pooled before being fed into the final classification layer \cite{hand}.
\iffalse
% \begin{figure*}[t]%
%     \centering
%     \includegraphics[width=12cm]{lstm_new_final.png}
%     \caption{ The graphical representation of one DS-LSTM cell}
%     \label{fig:2}%
% \end{figure*}
\fi
\subsubsection{CNN for Mel-Spectrograms}
\label{sssec: cnn}
As mentioned earlier, for each utterance $u_i$, we produce two mel-spectrograms with different time-frequency resolutions. We pass these two spectrograms into two independent 2D CNN blocks, each of which consist of two convolution and max-pooling layers. After both spectrograms go through the two convolution and max-pooling layers, they have a different number of time steps, one with $T_1$ and the other with $T_2$, where $T_1\approx2T_2$. Before passing both sequences into the DS-LSTM, we use an alignment procedure to ensure they have the same number of time steps, taking the average of adjacent time steps in the sequence of length $T_1$. After alignment, both sequences have the same number of time steps $T_3$, where $T_3\approx T_2$.
\subsubsection{Dual-Sequence LSTM}
\label{sssec: dslstm}
Following the alignment operation, we obtain two sequences of data, $X=\{x_1,\dots,x_{T_3}\}$ and $Y=\{y_1,\dots,y_{T_3}\}$, with the same number of time steps. Here, $X$ comes from mel-spectrogram $S_{1i}$, which records more information along the time axis, and $Y$ comes from mel-spectrogram $S_{2i}$, which records more information along the frequency axis. It is entirely conceivable that sequences $X$ and $Y$ will complement each other, and therefore it will be beneficial to process them through a recurrent network simultaneously.

\begin{figure}
    \centering
    \includegraphics[width=7.5cm]{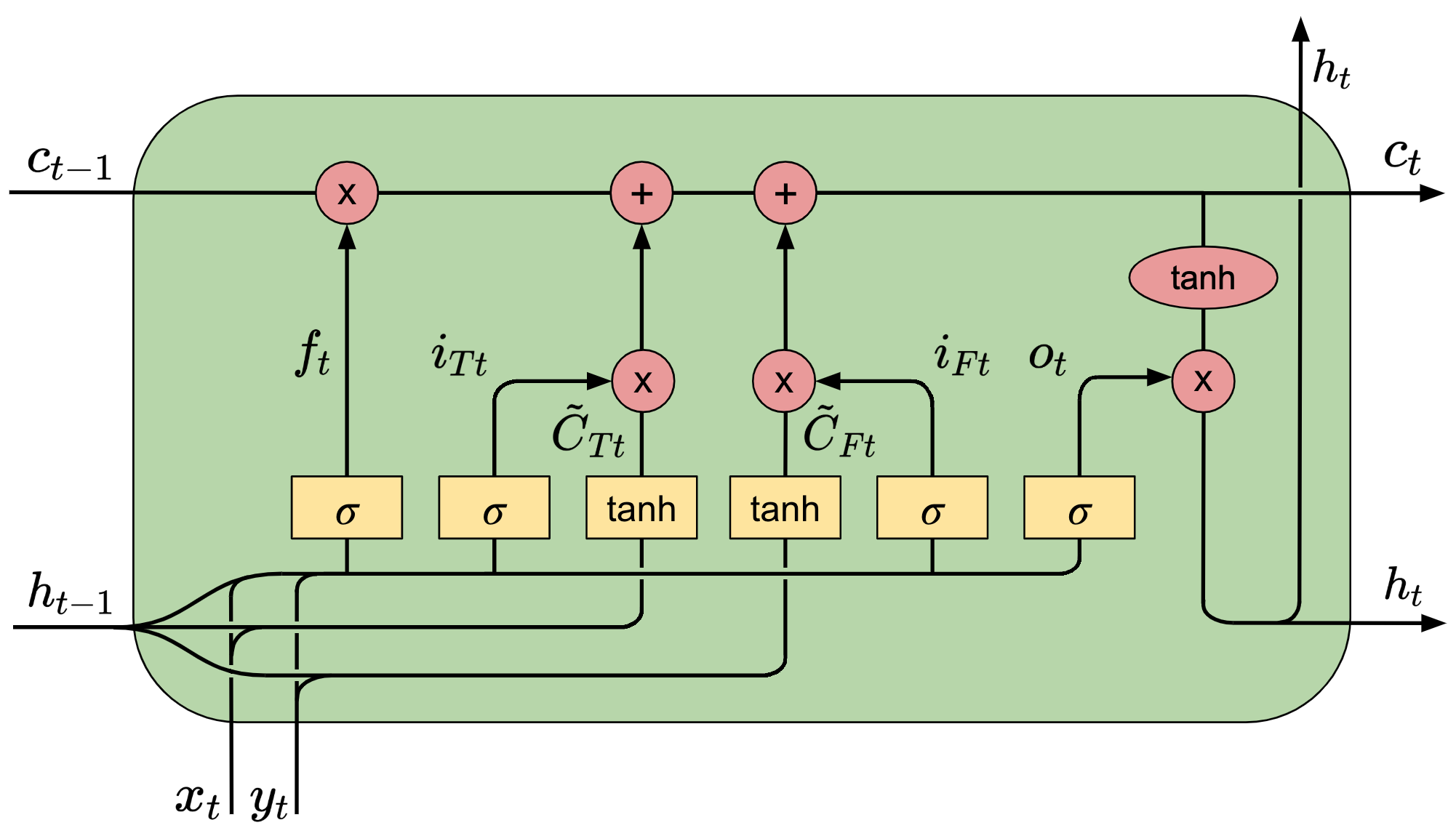}
    \caption{The graphical representation of one DS-LSTM cell}
    \label{fig:2}%
\end{figure}

As Figure~\ref{fig:2} indicates, we propose a Dual-Sequence LSTM (DS-LSTM) that can process two sequences of data simultaneously. Let $\odot$ denote the Hadamard product, $[a,b]$ the concatenation of vectors, $\sigma$ the sigmoid activation function, $\tanh$ the hyperbolic tangent activation function, and \text{rbn} the recurrent batch normalization layer, which keeps a separate running mean and variance for each time step \cite{rbn}.
\begin{equation}
f_{t}=\text{rbn}(\sigma(W_{f}[x_t,y_t,h_{t-1}]+b_{f}))
\end{equation}
\begin{equation}
i_{T_t}=\text{rbn}(\sigma(W_{i_T}[x_t,y_t,h_{t-1}]+b_{i_T}))
\end{equation}
\begin{equation}
i_{F_t}=\text{rbn}(\sigma(W_{i_F}[x_t,y_t,h_{t-1}]+b_{i_F}))
\end{equation}
\begin{equation}
o_{t}=\text{rbn}(\sigma(W_{o}[x_t,y_t,h_{t-1}]+b_{o}))
\end{equation}
\begin{equation}
\tilde{C_{T_t}}=\tanh{(W_{T}[x_t,h_{t-1}]+b_{T}})
\end{equation}
\begin{equation}
\tilde{C_{F_t}}=\tanh{(W_{F}[y_t,h_{t-1}]+b_{F}})
\end{equation}
\begin{equation}
C_t=f_t \odot C_{t-1} +i_{T_t}\odot \tilde{C_{T_t}}+i_{F_t} \odot \tilde{C_{F_t}}
\end{equation}
\begin{equation}
h_t=o_t \odot \tanh{(C_t)}
\end{equation}

After the execution of (8), $h_t$ is the hidden state for the next time step, but $h_t$ also goes through a batch normalization layer to be the input for the next layer of the DS-LSTM at time $t$.

While an LSTM is a four-gated RNN, the DS-LSTM is a six-gated RNN, with one extra input gate $i_{F_t}$ at (3) and one extra intermediate memory cell $\tilde{C_{F_t}}$ at (6). The two intermediate memory cells $\tilde{C_{T_t}}$ and $\tilde{C_{F_t}}$ are derived from $X$ and $Y$, respectively, with the intuition that $\tilde{C_{T_t}}$ will capture more information along the time axis, while $\tilde{C_{F_t}}$ will capture more information along the frequency axis. Empirical experiments suggest that the forget gate, two input gates, and output gate should incorporate the maximum amount of information, which is the concatenation of $x_t, y_t, $ and $h_{t-1}$. 

A recurrent batch normalization layer (rbn) is used to normalize the output of the forget gate, input gates, and output gate in order to speed up training and provide the model with a more robust regularization effect.

In summary, Section~\ref{sssec: lstm} describes the vanilla model $M_{\text{LSTM}}$. Sections~\ref{sssec: cnn} and~\ref{sssec: dslstm} describe the architecture for our proposed DS-LSTM model, denoted as $M_{\textrm{DS-LSTM}}$. Together, $M_{\text{LSTM}}$+$M_{\textrm{DS-LSTM}}$ describes our proposed Dual-Level model as illustrated in Figure~\ref{fig:1}.

\section{Experimental Setup and Results}
\subsection{Experimental Setup}
\label{ssec:ex}
For the CNN block used to process the mel-spectrograms, a 4$\times$4 kernel is used without padding, and the max pooling kernel is 2$\times$2 with a 2$\times$2 stride. For each layer of the CNN, the output channels are 64 and 16, respectively. All gate neural networks within LSTM and DS-LSTM have 200 hidden nodes. Each LSTM is single-directional with two layers. The weight and bias for the recurrent batch normalization parameters are initialized as 0.1 and 0, respectively, as suggested by the original paper \cite{rbn}. An Adam optimizer is used with the learning rate set at 0.0001.
\subsection{Baseline Methods}
\label{ssec:baseline}
Since several modifications are proposed, we create six baseline models that consist of various parts of the whole model in order to better evaluate the value of each modification.\\ \\
\textbf{Base 1}: $M_{\textrm{LSTM}}$, which is the LSTM-based model with the MFCC features.\\
\textbf{Base 2}: CNN+LSTM, whose inputs, $\{S_{11},\dots,S_{1n}\}$, are spectrograms with 256 FFT points. Inputs are passed through a CNN followed by an LSTM. Models such as these are developed in \cite{8682154} and \cite{Etienne2018CNNLSTMAF}.\\
\textbf{Base 3}: CNN+LSTM, whose inputs, $\{S_{21},\dots,S_{2n}\}$, are spectrograms with 512 FFT points. Inputs are passed through a CNN followed by an LSTM. Note that the architecture is the same as \textbf{Base 2}.\\
\textbf{Base 4}: A combination of models of \textbf{Base 2} and \textbf{Base 3}: $2\times (\textrm{CNN+LSTM})$, whose inputs are $\{S_{11},\dots,S_{1n}\}$ and $\{S_{21},\dots,S_{2n}\}$. In this model, two LSTMs process two sequences of mel-spectrograms separately, and their respective outputs are averaged to make final classifications. Note this is different from our proposed DS-LSTM, which processes these two sequences within a single DS-LSTM cell.\\
\textbf{Base 5}: A combination of models of \textbf{Base 1} and \textbf{Base 4}. \\
\textbf{Base 6}: A combination of models of \textbf{Base 1} and \textbf{Base 2}.\\ \\
In addition to the above six baseline models, we propose two models, $M_{\textrm{DS-LSTM}}$ and the dual-level model, $M_{\textrm{LSTM}}$+$M_{\textrm{DS-LSTM}}$. We compare these models with the baseline models, as well as four state-of-the-art models that use standard 5-fold cross-validation for evaluation.
\subsection{Results and Analysis}
\label{ssec: ra}
\definecolor{Gray}{gray}{0.9}
\begin{table}[H]
    \centering
    \begin{tabular}{|c|c|c|}
    \hline
        &Mean WA &Mean UA\\ \hline
        \rowcolor{Gray}
        {Base 1} = $M_{\textrm{LSTM}}$&64.7$\pm$1.4&65.5$\pm$1.7\\ 
        {Base 2} = CNN+LSTM&63.5$\pm$1.6&64.5$\pm$1.5\\ 
        \rowcolor{Gray}
        {Base 3} = CNN+LSTM&62.9$\pm$1.0&64.3$\pm$0.9\\ 
        {Base 4} = {Base 2} + {Base 3}&64.4$\pm$1.8&65.2$\pm$1.8\\ 
        \rowcolor{Gray}
        {Base 5} = {Base 1} + {Base 4}&68.3$\pm$1.3&69.3$\pm$1.2\\ 
        {Base 6} = {Base 1} + {Base 2}&68.5$\pm$0.8&68.9$\pm$1.2\\\hline
        \rowcolor{Gray}
        D. Dai et. al (2019) \cite{centerloss} &65.4&66.9\\
        S. Mao et. al (2019) \cite{markov}&65.9&66.9\\
        \rowcolor{Gray}
        R. Li et. al (2019) \cite{8682154}&---&67.4\\ 
        S. Yoon et. al (2018) \cite{multi2} *&71.8$\pm$1.9&---\\ \hline
        \rowcolor{Gray}
        \textbf{Proposed} \boldsymbol{$M_{\textrm{DS-LSTM}}$}&69.4$\pm$0.6&69.5$\pm$1.1\\
        \textbf{Proposed} \boldsymbol{$M_{\textrm{LSTM}}$}+\boldsymbol{$M_{\textrm{DS-LSTM}}$}&72.7$\pm$0.7&73.3$\pm$0.8\\ \hline

    \end{tabular}
    \caption{\small{Mean WA and Mean UA are the average of weighted accuracy and unweighted accuracy, respectively, for 5-fold cross validation. Most results are reported with one standard deviation. \\ * indicates the model uses textual information.}}
    \label{table:stats}
\end{table}
Table~\ref{table:stats} indicates our proposed model $M_{\textrm{LSTM}}$+$M_{\textrm{DS-LSTM}}$ outperforms all baseline models by at least 4.2\% in mean weighted accuracy, and by at least 4.0\% in mean unweighted accuracy. It also outperforms state-of-the-art unimodal SER models \cite{centerloss,markov,8682154} by at least 6.8\% in mean weighted accuracy and 5.9\% in mean unweighted accuracy. Although multimodal SER models typically have a higher accuracy due to its access to both audio data and textual data, we see that our proposed model achieves comparable performance with \cite{multi2} in mean weighted accuracy.

Before further investigating the effectiveness of each integrated part of the proposed dual-level model $M_{\textrm{LSTM}}$+$M_{\textrm{DS-LSTM}}$, we note that Base 1$\sim$3 and 6 have less parameters than our proposed models. However, we have verified that simply adding more nodes or layers to these models does not make any empirical difference in its predictive power, which suggests that these aforementioned baseline models have already reached their full potential. Therefore, we can objectively compare these models.

Both Base 2 and Base 3 take a single sequence of mel-spectrograms, and both perform slightly worse than Base 1, which only uses MFCC features. This supports the claim that mel-spectrograms are harder to learn than MFCC features. Base 4 is a naive combination of Base 2 and Base 3, and because the two LSTMs in Base 4 do not interact with each other, the complementary information between these two sequences of mel-spectrograms is not fully explored; therefore, Base 4 is also slightly worse than Base 1. Base 5 and Base 6 are both dual-level models that consider both MFCC features and mel-spectrograms, and they both outperform Base 1$\sim$4, demonstrating the effectiveness of the dual-level model. 

More importantly, we observe that the proposed $M_{\textrm{DS-LSTM}}$ significantly outperforms Base 1$\sim$4. Comparing $M_{\textrm{DS-LSTM}}$ with Base 4, we see that when two separate LSTMs are replaced by the DS-LSTM, which has only six neural networks in its cell instead of eight neural networks in two LSTMs together, the weighted accuracy increases by 5\% and the parameters are reduced by 25\%. This shows that the DS-LSTM is a successful upgrade from two separate LSTMs. When we consider the dual-level model $M_{\textrm{LSTM}}$+$M_{\textrm{DS-LSTM}}$, it outperforms all baseline methods significantly. 

\section{Conclusion}
In this paper, we have demonstrated the effectiveness of combining MFCC features and mel-spectrograms produced from audio signals for emotion recognition. Furthermore, we introduced a novel LSTM architecture, denoted as DS-LSTM, which can process two mel-spectrograms simultaneously. We also outlined several modifications to the data preprocessing step. Our proposed model significantly outperforms baseline models and current state-of-the-art unimodal models on the IEMOCAP dataset, and is comparable with multimodal models, showing that unimodal models, which only rely on audio signals, have not reached their full potential.
\section{Acknowledgements}
We are grateful for the assistance of Mr. Reza Soleimani and the support of Professor Robert Calderbank from the Rhodes Information Initiative at Duke University.
%Since it is beneficial to look at spectrograms from multiple time-frequency resolutions, one possible future direction is to enable the DS-LSTM to process multiple sequences of data in a fast and memory-efficient way without overfitting. We also would like to explore different options of increasing speaker variability, since we consider one of the primary challenges of classifying utterances on the IEMOCAP dataset to be its lack of speaker variability.

\newpage
\bibliography{cite}
\bibliographystyle{IEEEbib}
\end{document}